%% file: paper.tex
\documentclass[a4paper,11pt]{article}
\usepackage[zerostyle=e]{newtxtt}
\usepackage{pos}
\input{include}

\title{The Equation of State of QCD up to very high temperatures}
\author*[a,b]{Matteo Bresciani}
\author*[a]{Michele Pepe}
\author[a,b]{Mattia Dalla Brida}
\author[a,b]{Leonardo Giusti}

\affiliation[a]{University of Milano-Bicocca, Piazza della Scienza 3, Milan, I-20126, Italy}
\affiliation[b]{INFN Milano-Bicocca, Piazza della Scienza 3, Milan, I-20126, Italy}

\emailAdd{m.bresciani9@campus.unimib.it}
\emailAdd{michele.pepe@mib.infn.it}
\emailAdd{mattia.dallabrida@unimib.it}
\emailAdd{leonardo.giusti@unimib.it}

\abstract{
We present the non-perturbative computation of the entropy density in QCD for temperatures 
ranging from 3 GeV up to the electro-weak scale, using $N_f=3$ flavours of massless O$(a)$-improved Wilson fermions. 
We adopt a new strategy designed to be computationally efficient and based on formulating thermal QCD in a moving reference frame, where the
fields satisfy shifted boundary conditions in the temporal direction and periodic boundary conditions along the spatial ones.  
In this setup the entropy density can be computed as the derivative of the free-energy density with respect to the shift parameter. 
For each physical temperature, we perform Monte Carlo simulations at four values of the lattice spacing in order to extrapolate the
numerical data of the entropy density to the continuum limit. We achieve a final accuracy of approximatively $0.5$-$1.0\%$ and our results are
compared with predictions from high-temperature perturbation theory. 
}

\FullConference{The 41st International Symposium on Lattice Field Theory (LATTICE2024)\\
 28 July - 3 August 2024\\
Liverpool, UK\\}

\begin{document}
\maketitle

\section{Introduction}
There is significant interest from particle physics to cosmology in understanding the thermodynamic properties of the high temperature
regime of QCD. In particular, the Equation of State (EoS) at all temperatures up to the electro-weak scale is crucial theoretical input
for cosmological models that describe the early evolution of the Universe. For instance, it is believed that the EoS played a key role in
shaping the spectrum of primordial gravitational waves and determining the abundance of several dark matter candidates, such as WIMPs and/or
axions~\cite{Saikawa:2018rcs,Saikawa:2020swg}. In particle physics, the EoS is required in the analysis of data coming from heavy-ion
colliders and for the study of QCD under extreme conditions. 

The EoS has been determined from non-perturbative lattice calculations up to temperatures of about 1 GeV in various staggered quark
setups~\cite{Borsanyi:2013bia,HotQCD:2014kol,Bazavov:2017dsy,Borsanyi:2016ksw}, with a precision of a few percent. Beyond this range, it is
known only from perturbative computations resummed using thermal effective theories of QCD~\cite{Appelquist:1981vg,Kajantie:2002wa} 
and from hard-thermal-loop perturbation theory~\cite{Andersen:2010ct,Andersen:2011sf}, which rely 
on small values of the running coupling constant holding at asymptotically high temperatures. However, the perturbative expansion suffers from
poor convergence, leading to an unsatisfactory description of the thermodynamic features of quarks and gluons up to the electro-weak
scale~\cite{Giusti:2016iqr,DallaBrida:2021ddx,Giusti:2024ohu}.

In these Proceedings we present the results of Ref.~\cite{Bresciani:2025vxw} on the non-perturbative determination of 
the QCD EoS with a precision of about $0.5$-$1.0\%$ in the unexplored
range of temperatures between $3$ and $165$ GeV. We carried out the calculation by numerical simulations with $\Nf=3$ massless flavours of
O$(a)$-improved Wilson fermions. This result is achieved through a new strategy based on two key novelties. 
First, instead of using a hadronic scheme, we determine the lines of constant physics by imposing that a
renormalized coupling, defined non-perturbatively in a finite volume scheme, assumes a prescribed value~\cite{DallaBrida:2021ddx}. 
Second, we consider the setup of thermal QCD in
a moving reference frame, which allows for the direct computation of the entropy density by exploiting the (Euclidean) Lorentz invariance of the
theory~\cite{Giusti:2012yj}. In particular, this approach prevents the need of the zero-temperature subtraction usually required for the
definition of thermodynamic potentials. 
The combination of these elements allows us to overcome the window problem that arises when simulating a high-energy scale and a
low-energy one at the same lattice spacing: in our case, the temperature and the hadronic scale, respectively. This strategy makes the
high-temperature regime accessible to non-perturbative lattice studies with controlled systematic effects and affordable computational
effort. It has already proved successful for the non-perturbative determination of the EoS of SU$(3)$ Yang-Mills
theory~\cite{Giusti:2016iqr} and, more recently, of the screening spectrum of QCD~\cite{DallaBrida:2021ddx,Giusti:2024ohu} up to very high
temperatures.

\section{EoS from QCD in a moving reference frame}
\label{sec:EoS_from_QCD_with_shifted_boundary_conditions}
The SO$(4)$ Euclidean invariance of QCD in the continuum allows us to define the thermal theory in a moving reference frame.
In the path integral formalism, this corresponds to introducing a spatial shift $\vxi=(\xi_1,\xi_2,\xi_3)$ when closing the boundary
conditions of the gauge field, $A_\mu$, and of the fermionic fields, $\psi$ and $\psibar$, along the compact direction of length
$L_0$~\cite{Giusti:2012yj,DallaBrida:2020gux}: 
\begin{equation}
\begin{aligned}
    A_\mu(x_0+L_0, \bs{x}&) = A_\mu(x_0, \bs{x}-L_0\vxi )\,, \\
    \psi(x_0+L_0, \bs{x}) &= -\psi(x_0, \bs{x}-L_0\vxi )\,, \\
    \psibar(x_0+L_0, \bs{x}) &= -\psibar(x_0, \bs{x}-L_0\vxi )\,.
\end{aligned}
\label{eq:shBCs}
\end{equation}
Periodic boundary conditions can be set in the other directions, taken of length $L$. In the presence of shifted boundary
conditions~\cite{Giusti:2012yj}, the thermal system has spatial size $V_s=L^3/\sqrt{1+\vxi^2}$ and it is at the temperature $T^{-1}=L_0
\sqrt{1+\vxi^2}$. The free-energy density of the thermal system is related as usual to the partition function $\cZ$,  
\begin{equation}
    f_\vxi = -\frac{T}{V_s}\ln\cZ = -\frac{1}{L_0 L^3}\ln\cZ\,, \quad \cZ = \int DAD\psibar D\psi\, e^{-S}\,,
\end{equation}
where $S$ is the QCD action.
Taking into account that, in the framework of shifted boundary conditions, the temperature can be changed by varying $\vxi$ at fixed $L_0$ and
that the pressure $p=-f_\vxi$, we obtain the following equation~\cite{Giusti:2012yj}
\begin{equation}
    \frac{s}{T^3} = \frac{1+\vxi^2}{\xi_k}\frac{1}{T^4}\frac{\partial f_\vxi}{\partial \xi_k}
    \label{eq:entropy_cont}
\end{equation}
which relates the entropy density $s=dp/dT$ to the derivative of the free-energy density $f_\vxi$ with respect to the shift.
The power divergence resulting from the mixing of the free energy with the identity is removed by the derivative in the
shift and the entropy defined in Eq.~\eqref{eq:entropy_cont} is directly a physical quantity. The pressure $p(T)$ can then be obtained by
integrating the entropy in the temperature, while the energy density $e(T)$ follows from the relation $Ts=e+p$.

\section{Lattice setup and renormalization}
\begin{table}
	\centering
	\begin{tabular}{ccc}
		\hline
		$T$ &  $\bar{g}^2_{\rm SF}(\mu=T\sqrt{2})$ & $T$ (GeV) \\
		\hline
		$T_0$ &  1.01636 &  164.6(5.6) \\
		$T_1$ &  1.11000 &  82.3(2.8)  \\
		$T_2$ &  1.18446 &  51.4(1.7)  \\
		$T_3$ &  1.26569 &  32.8(1.0)  \\
		$T_4$ &  1.3627  &  20.63(63)  \\   
		$T_5$ &  1.4808  &  12.77(37)  \\
		$T_6$ &  1.6173  &  8.03(22)   \\
		$T_7$ &  1.7943  &  4.91(13)   \\
		$T_8$ &  2.0120  &  3.040(78)  \\
		\hline
	\end{tabular}
	\caption{Values of the Schr\"odinger functional coupling in $\Nf=3$ QCD with Wilson action
    used to renormalize the bare parameters at the physical temperatures considered in this work, 
    reported in the last column.}
  \label{tab:T0T8GeV}	
\end{table}
The QCD action on the lattice, $S=S_G+S_F$, is given by the sum of a gluonic and a fermionic component. The gluonic part, $S_G$,
involves only the gauge field and we consider the Wilson plaquette action 
\begin{equation}
    S_G = \frac{6}{g_0^2}\sum_x\sum_{\mu<\nu}\left[ 1 - \frac{1}{3}{\rm Tr}\left\{U_{\mu\nu}(x)\right\}\right]\,,
    \label{eq:SG}
\end{equation}
where the trace is over the colour index, $g_0$ is the bare gauge coupling and $\mu,\nu=0,1,2,3$. The plaquette field $U_{\mu\nu}$ is
defined as 
\begin{equation} 
    U_{\mu\nu}(x) = U_\mu(x) U_\nu(x+a\hat\mu)U_\mu^\dagger(x+a\hat\nu)U_\nu^\dagger(x)
    \label{eq:plaquette}
\end{equation}
where $U_\mu(x)\in$ SU$(3)$ is the link field. The fermionic part $S_F$ reads 
\begin{equation}
    S_F = a^4 \sum_x \psibar(x)\left(D+m_0\right)\psi(x)\,,
    \label{eq:SF}
\end{equation}
where $D$ is the O$(a)$-improved Wilson-Dirac operator and $m_0$ is the bare quark mass.
On the lattice, the link field and the fermionic fields satisfy shifted boundary conditions analogous to the ones
in Eq.~\eqref{eq:shBCs}, with the field $A_\mu$ replaced by $U_\mu$. 
The fields are periodic along the spatial directions.

In QCD with $N_f=3$, the non-perturbative running of the Schr\"odinger functional coupling $\bar{g}^2_{\rm SF}$
is known precisely in the continuum limit~\cite{Campos:2018ahf,DallaBrida:2018rfy,Bruno:2017gxd}.
Building on this result, we determine the lines of constant physics at a given temperature $T$ by fixing
the value of the renormalized coupling at finite lattice spacing to match its value in the continuum at a scale $\mu=T\sqrt{2}$ :
\begin{equation}
    \bar{g}^2_{\rm SF}(g_0^2, a\mu) = \bar{g}^2_{\rm SF}(\mu)\,, \quad a\mu\ll 1\,.
    \label{eq:SFrencond}
\end{equation}
This condition fixes the dependence of the bare coupling $g_0$ on
the lattice spacing, for values of $a$ at which the scale $\mu$ and therefore the temperature can
be easily accommodated. As a consequence, each temperature can be simulated at several
lattice resolutions and the continuum limit can be taken with confidence.
In this study we have considered 9 values of temperature between $3$ GeV and $165$ GeV, as reported in Table~\ref{tab:T0T8GeV}.
The critical mass $m_{\rm cr}$ at given bare parameters is then defined by requiring the PCAC mass to vanish 
in the Schr\"odinger functional setup. We refer to Appendix B and to Table 4 of~\cite{DallaBrida:2021ddx} for the technical details and
for the resulting lines of constant physics.

\section{Entropy density on the lattice}

As discussed in Section~\ref{sec:EoS_from_QCD_with_shifted_boundary_conditions}, the entropy density defined in Eq.~\eqref{eq:entropy_cont}
is directly a physical quantity. This allows us to discretize the equation on the lattice and to determine the entropy density by
extrapolating to the continuum limit the measurements obtained from Monte Carlo simulations performed at various lattice spacings, all
corresponding to the same fixed physical temperature $T$. Thus, at given values of $L_0/a$ and $g_0^2$, we have
\begin{equation}
    \frac{s}{T^3} = \frac{1+\vxi^2}{\xi_k}\frac{1}{T^4}
    \frac{\Delta f_\vxi}{\Delta\xi_k}\,,
    \label{eq:entropy_lattice}
\end{equation}
where 
\begin{equation}
    \frac{\Delta f_\vxi}{\Delta\xi_k} = 
    \frac{L_0}{4a}\left(f_{\vxi+\frac{2a}{L_0}\hat{k}} - f_{\vxi-\frac{2a}{L_0}\hat{k}}\right)
    \label{eq:f_diff_discrete_shift}
\end{equation}
is the 2-point discrete symmetric derivative of the free-energy with respect to the $k$-th component 
of the shift
\footnote{We use the two-point discrete derivative due to constraints related to the simulation software.}.
For computational efficiency, at fixed $L_0/a$ and $g_0^2$ the discrete derivative of the free-energy density 
is conveniently decomposed into two contributions,
\begin{equation}
    \frac{\Delta f_\vxi}{\Delta\xi_k} = 
    \frac{\Delta (f_\vxi - f_\vxi^\infty)}{\Delta\xi_k} + 
    \frac{\Delta f_\vxi^\infty}{\Delta\xi_k}\,,
    \label{eq:fYM_diff}
  \end{equation}
where $f_\vxi^\infty$ is the free-energy density of QCD with infinitely massive quarks, i.e. in the static limit.
The first term can be rewritten as
\begin{equation}
   \frac{\Delta (f_\vxi - f_\vxi^\infty)}{\Delta\xi_k} =  
   - \frac{\Delta}{\Delta\xi_k} \int_0^\infty dm_q\frac{\partial f_\vxi^{m_q}}{\partial m_q}
   = -\int_0^\infty dm_q \frac{\Delta\corrno{\psibar\psi}_\vxi^{m_q}}{\Delta\xi_k}\,,
   \label{eq:Df_quark}
\end{equation}
where $m_q = m_0 - m_{\rm cr}(g_0^2, L_0)$ is the bare subtracted quark mass. 
The discrete derivative of the chiral condensate $\corrno{\psibar\psi}_\vxi$ with respect to the shift is defined similarly to
Eq.~\eqref{eq:f_diff_discrete_shift}. The second term in Eq.~\eqref{eq:fYM_diff} is evaluated in the static quark limit of QCD. 
We rewrite it as follows, 
\begin{equation}
    \frac{\Delta f_\vxi^\infty}{\Delta\xi_k} = 
    \frac{\Delta f_\vxi^{\infty,(0)}}{\Delta\xi_k}
    + g_0^2\, \frac{\Delta f_\vxi^{\infty, (1)}}{\Delta\xi_k}
    - \int_0^{g_0^2} du \left(\frac{1}{u}\left.\frac{\Delta\corrno{S_G}_\vxi^\infty}{\Delta\xi_k}\right|_{g_0^2=u}
    +\frac{\Delta f_\vxi^{\infty, (1)}}{\Delta\xi_k}\right)\,,
    \label{eq:Df_gauge}
\end{equation}
where $f_\vxi^{\infty, (0)}$ and $f_\vxi^{\infty, (1)}$ represent, respectively, the tree-level and one-loop coefficients 
of the perturbative expansion on the lattice of the free-energy density in pure gauge theory.

In the following we discuss the numerical evaluation on the lattice of the two contributions
in Eqs.~\eqref{eq:Df_quark} and~\eqref{eq:Df_gauge}. 
We consider four resolutions $L_0/a=4,6,8,10$ of the compact direction, while the three
spatial directions are taken to be of the same size, equal to $L/a = 144$.
Finite volume effects are exponentially suppressed as $e^{-M_{\rm gap}L}$ for
$L\to\infty$, where $M_{\rm gap}$ is the mass gap of the theory. 
In the high-temperature regime that we are investigating, we have $M_{\rm gap}\propto T$ with a coefficient close to
unity~\cite{Giusti:2012yj,DallaBrida:2021ddx},  
and therefore we expect finite size effects to be well below the statistical accuracy of our numerical results.
We choose the shift vector $\vxi=(1,0,0)$ as it turned out to lead to milder discretization effects
in perturbation theory and in other studies employing shifted 
boundary conditions~\cite{Giusti:2015daa,DallaBrida:2021ddx,Bresciani:2022lqc}.

\begin{figure}
     \centering
     \begin{subfigure}{0.48\textwidth}
         \centering
         \includegraphics[width=\textwidth]{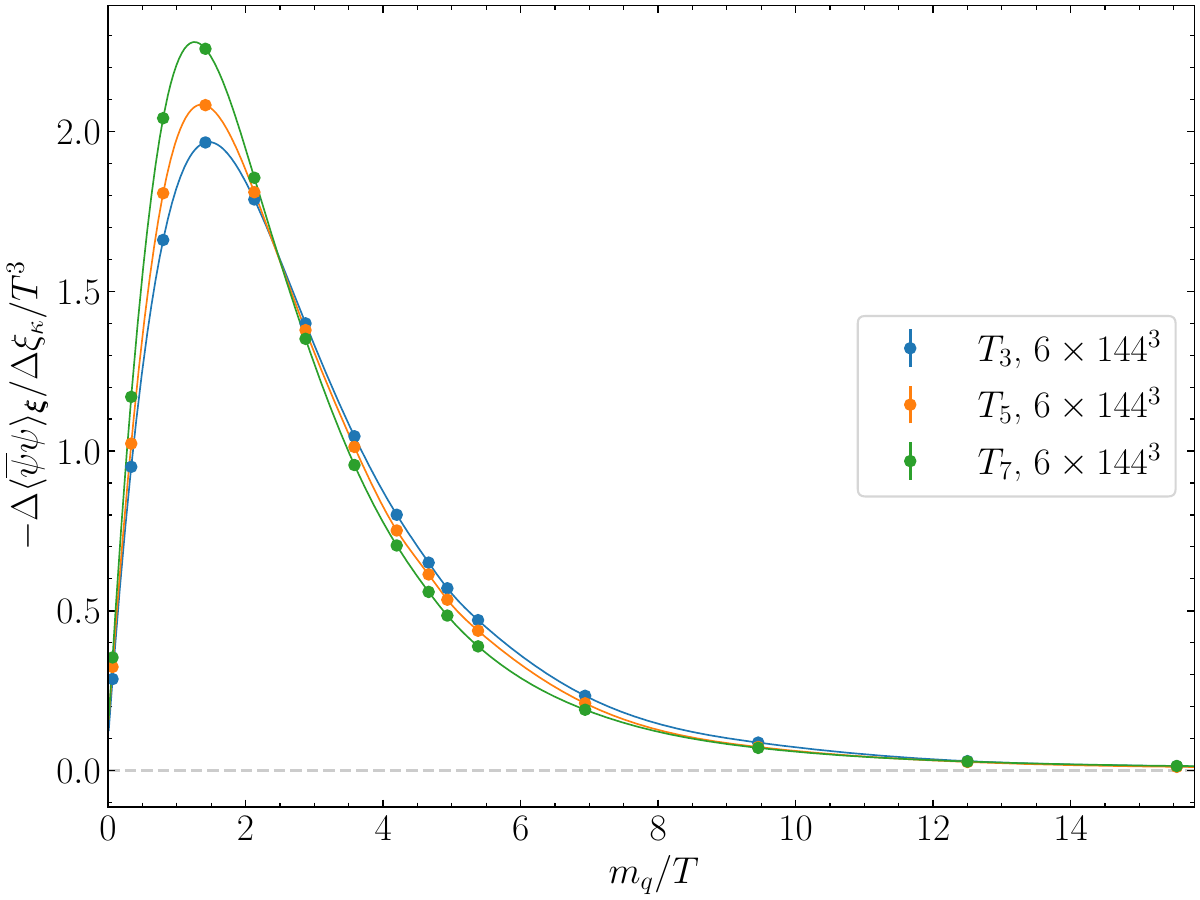}
     \end{subfigure}%
     \hspace{0.4cm}
     \begin{subfigure}{0.48\textwidth}
         \centering
         \includegraphics[width=\textwidth]{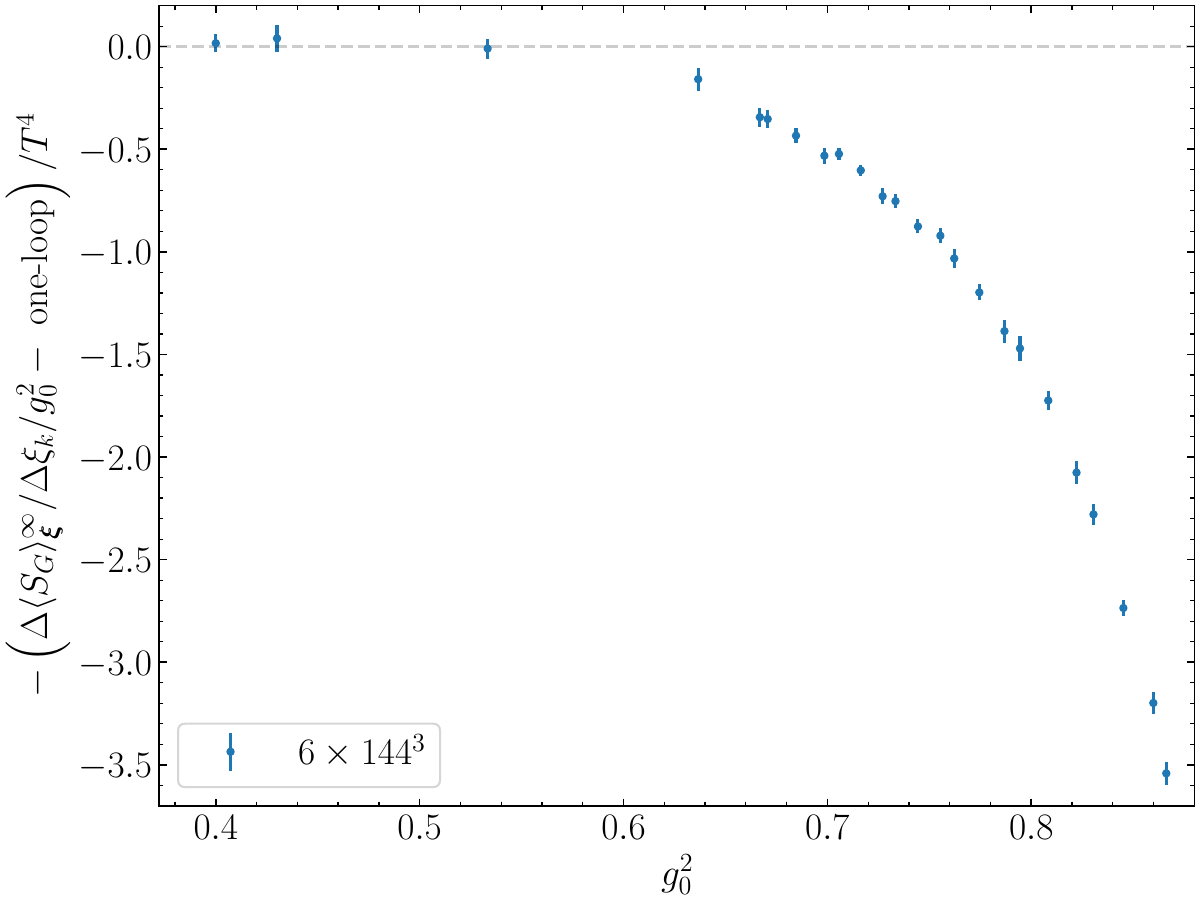}
     \end{subfigure}
     \caption{Left: Derivative in the shift of the chiral condensate as a
     function of $m_q/T$ at some selected bare parameters. 
     Points have been interpolated with a cubic spline to guide the eye.
     Error bars are smaller than the markers.
     Right: Derivative in the shift of the pure gauge action as a function 
     of $g_0^2$ for $L_0/a = 6$. 
     For convenience we subtracted from the data the result at one-loop order 
     in lattice perturbation theory.
     }
     \label{fig:integrand_functions}
\end{figure}

\subsection{Integration in the bare mass}
In this subsection we discuss the method we have used to compute the integral in the bare subtracted 
quark mass appearing in Eq.~\eqref{eq:Df_quark}. At given values of $L_0/a$ and $g_0^2$, we split the integration interval
into three  domains. The first one, corresponding to $m_q/T\in[0, 5]$, gives the largest contribution to the integral and it is computed
with a 10-point Gaussian quadrature. The second interval, where we go up to $m_q/T=20$ ($m_q/T=35$ for $L_0/a=4$), is computed with a 
7-point Gaussian quadrature and gives about $20\%$ of the result. Finally, the last interval comes from the asymptotic tail
$m_q/T\to\infty$ and contributes to the total by two standard deviations at most. It is estimated with a 3-point Gaussian quadrature, 
after a change of variable to the hopping parameter 
$\kappa = 1/(2am_0+8)$ which leads to a compact integration interval. This combination of Gaussian quadratures has been chosen and optimized
by using perturbation theory as guidance so as to keep the systematic error coming from the numerical integration well below the statistical
accuracy of the numerical results. 

The computation of the discrete derivative in the integrand of Eq.~\eqref{eq:Df_quark} requires performing two Monte Carlo simulations to
measure the chiral condensate at the two shifts $\vxi=(1\pm 2a/L_0, 0, 0)$. By carrying out this calculation for all the bare mass values
determined by the Gaussian quadratures, we obtain a total of 40 independent simulations for each given value of $L_0/a$ and $g_0^2$.
The left panel of Figure~\ref{fig:integrand_functions} shows the resulting integrand function
for a selection of bare parameters. The final accuracy on the integral ranges from a few permille for $L_0/a=4$ to about $1\%$ for  $L_0/a=10$.

\subsection{Integration in the bare coupling}
The computation of the integral in Eq.~\eqref{eq:Df_gauge} is carried out in SU(3) Yang-Mills theory. Although this contribution has 
larger variance compared to the one in Eq.~\eqref{eq:Df_quark}, pure gauge simulations are significantly less computationally demanding and
high precision can be reached with a moderate effort.
For a given value $L_0/a$, let $g_0^2|_{T_i}$ be the value of the bare coupling determined from the lines of constant physics 
at temperature $T_i$ with $i=0,1,...,8$. The observable we have to integrate is $\Delta\corrno{S_G}_\vxi^\infty/\Delta\vxi_k$ and we split
the integration interval $[0,g_0^2|_{T_i}]$ into several parts combining different integration schemes.

In the domain $g_0^2\in[0,\,6/15]$ the integral is computed with a 2-point trapezoidal rule (3-point Simpson rule for $L_0/a=4$).
The interval $g_0^2\in[6/15,\,6/9]$ is integrated with a 3-point Gaussian quadrature.
For the resolutions $L_0/a=4,6$, the domain $g_0^2\in[6/9,\,g_0^2|_{T_0}]$ is integrated
with a 3-point Gaussian quadrature for $L_0/a=4$ and with the midpoint rule for $L_0/a=6$.
The interval $g_0^2\in[6/9,\,g_0^2|_{T_1}]$ is computed with a 3-point Gaussian quadrature.
At the lower temperatures $T_i$, $i>1$, the value of the integral is obtained by adding to 
the $T_{i-1}$ result the integral in the segment $g_0^2\in[g_0^2|_{T_{i-1}},\,g_0^2|_{T_i}]$,
estimated with a 3-point Gaussian quadrature for $i=2,3,4,5,6$, and a 5-point Gaussian quadrature for $i=7,8$.

Similarly to the fermionic case, the computation of the discrete derivative $\Delta\corrno{S_G}_\vxi^\infty/\Delta\vxi_k$ requires
performing two Monte Carlo simulations to measure $\corrno{S_G}^\infty_\vxi$ at the two shifts $\vxi=(1\pm 2a/L_0, 0, 0)$.
This leads to a total of 74 independent simulations for $L_0/a=4$, 68 for $L_0/a=6$ and 66 for $L_0/a=8,10$.
The pure gauge ensembles for this computation have been generated in Monte Carlo simulations 
where the basic sweep is a combination of heatbath and overrelaxation~\cite{Adler:1987ce} updates of the link variables, 
using the Cabibbo–Marinari scheme~\cite{Cabibbo:1982zn}.
The right panel of Figure~\ref{fig:integrand_functions} shows a representative case for the integrand function in $g_0^2$.
At each temperature, the accuracy on the integral is $\sim0.5\%$ for $L_0/a=4,6$, $\sim1\%$ for $L_0/a=8$ and $\sim1.5\%$ for $L_0/a=10$.

\subsection{Continuum limit}

\begin{figure}
     \centering
     \begin{subfigure}{0.48\textwidth}
         \centering
         \includegraphics[width=\textwidth]{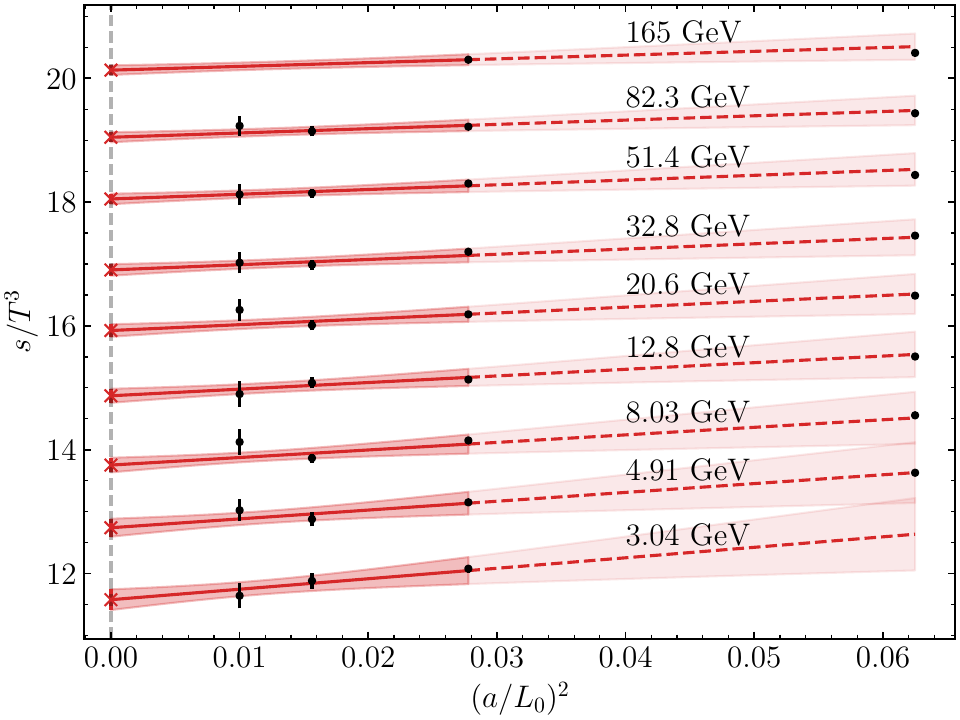}
     \end{subfigure}%
     \hspace{0.2cm}
     \begin{subfigure}{0.48\textwidth}
         \centering
         \includegraphics[width=\textwidth]{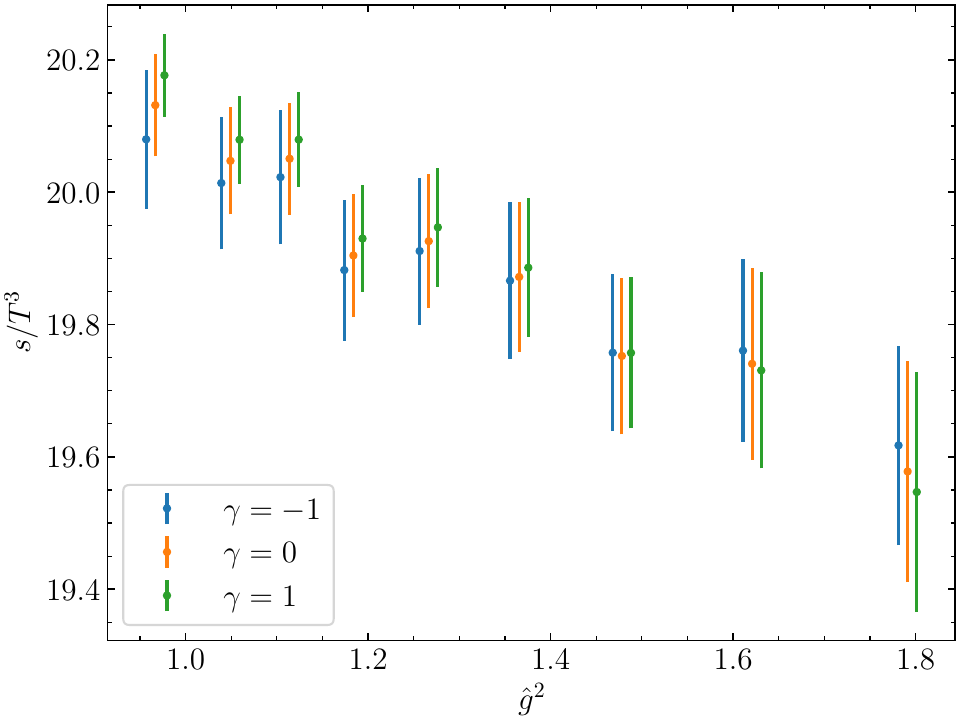}
     \end{subfigure}
     \caption{Left: The black points are the values of $s/T^3$ at fixed lattice spacing
     $a/L_0$ and given temperature. Points at temperature $T_n$ have been shifted 
     downward by $n$ for better readability. The red bands represent our best fit to
     the continuum limit.
     Right: Effect of logarithmic corrections on the continuum extrapolated values
     of the best fit, see main text for details.}
     \label{fig:clim}
\end{figure}

In this subsection we discuss the extrapolation to the continuum limit of the results for the entropy density we have obtained by Monte
Carlo simulations. At all orders in the lattice spacing, discretization effects of order O$(g^0)$ and O$(g^2)$ can be removed by improving 
the observable at one-loop order. This can be accomplished, for instance, considering 
\begin{equation}
    s(L_0/a, g_0^2) \to 
    s(L_0/a, g_0^2)\,\frac{s_0 + s_2 \cdot 
    \left(\frac{g}{2\pi}\right)^2}{s_0(L_0/a) + s_2(L_0/a) 
    \cdot \left(\frac{g}{2\pi}\right)^2}
\end{equation}
where $g^2\equiv\bar{g}_{\rm SF}^2(1/L_0)$ while $s_0(L_0/a)$ and $s_2(L_0/a)$ are the tree-level and the one-loop coefficients computed in
lattice perturbation theory in the thermodynamic limit. Their values for the relevant $L_0/a$ are listed in Table~\ref{tab:entropy_lpt},
together with the results in the continuum limit. After performing this one-loop improvement of the lattice results, the leading 
discretization effects are expected to be of O($g^3a^2$). On general grounds, at finite temperature, we expect also odd powers 
in the coupling to be present.

\begin{table}
    \centering
    \begin{tabular}{cccccc}
    \hline
    $L_0/a$ & $4$ & $6$ & $8$ & $10$ & $\infty$ \\
    \hline
    $s_0$ & $4.719$ & $3.438$ & $3.145$ & $3.059$ & $2.969$ \\
    $s_2$ &  $-23.465$ & $-13.085$ & $-9.955$ & $-9.089$ & $-8.438$ \\
    \hline
    \end{tabular}
    \caption{Coefficients for the one-loop improvement of the lattice entropy density.}
    \label{tab:entropy_lpt}
\end{table}

We computed the entropy density in the continuum by extrapolating the one-loop improved lattice entropy density using a combined fit of the
data collected at the various temperatures. We have considered the following fit function,
\begin{equation}
    s(T_i, a/L_0)/T_i^3 = 
    s(T_i)/T_i^3 + d_2 \left(\frac{a}{L_0}\right)^2 g_i^3 
    + d_3 \left(\frac{a}{L_0}\right)^3 g_i^3
    \label{eq:fit_type}
\end{equation}
where $i=0,\ldots,8$ and $g_i\equiv\bar{g}_{\rm SF}(\sqrt{2}T_i)$.
The fit parameters $s(T_i)$ represent the continuum values at fixed temperature $T_i$, and the coefficients $d_2,d_3$ parametrize the
discretization effects. We first fitted data with $L_0/a=4,6,8,10$ by setting either $d_3=0$ or $d_3\neq0$.
While both fits have $\chi^2/\chi^2_{\rm exp}\approx 1$, with $\chi_{\rm exp}$ defined as in Ref.~\cite{Bruno:2022mfy}, the former gives
values for $s(T_i)$ with errors $3$-$4$ times smaller.
Although compatible within the larger errors of the second fit,
the extrapolated central values of the first fit are systematically higher.

A quadratic fit of the lattice artifacts ($d_3 = 0$), excluding the data at $L_0/a = 4$, yields $\chi^2/\chi^2_{\rm exp} \approx 0.7$. This
fit provides estimates of the continuum values $s(T_i)$ that are in good agreement with those obtained from the full dataset and the $d_3
\neq 0$ fit, both in terms of central values and error size. This suggests that the data at $L_0/a=4$ are likely affected by 
discretization effects of higher order than $a^2$. Thus, we used the data at the coarsest lattice spacing only to estimate 
the size of the O($a^3$) contributions and included these as a systematic error of the points at finer lattice spacing. 

More specifically, we used the value of $d_3$ obtained from the fit of the full dataset as an estimate of the systematic uncertainty and we
added it in quadrature to the statistical errors, $\sigma(T_i, a/L_0)$, of the normalized entropy density at given temperature and lattice
spacing
\begin{equation}
   \sigma^2(T_i, a/L_0) \to \sigma^2(T_i, a/L_0) + 
   \left[d_3(a/L_0)^3 g_i^3\right]^2\,.
   \label{eq:inflation}
\end{equation}
This quantity receives contributions from the statistical variance of the
Monte Carlo ensembles, from the uncertainties related to the definition 
of the lines of constant physics~\cite{DallaBrida:2018rfy,DallaBrida:2021ddx}
(although negligible with respect to the total), 
and from the correlations introduced by the integration in the bare 
coupling, see Eq.~\eqref{eq:Df_gauge}.
All the correlations have been properly taken into account using the tools 
of Refs.~\cite{Joswig:2022qfe,Ramos:2018vgu}.

The final best fit is the one considering data with $L_0/a>4$, $d_3=0$ in the 
fit ansatz Eq.~\eqref{eq:fit_type}, and errors as in Eq.~\eqref{eq:inflation}
for the definition of the weights in the $\chi^2$-function minimized by the fit.
The resulting values for $s(T_i)/T_i^3$ are reported in 
Table~\ref{tab:entropy_cont}, and have a relative error of $0.5$-$1.0\%$.
\begin{table}
    \centering
	\begin{tabular}{ccc}
		\hline
		$T$ & $T$ (GeV) & $s/T^3$ \\
		\hline
		$T_0$ &  165(6) & 20.13(8) \\
		$T_1$ &  82.3(2.8)  & 20.05(8)  \\
		$T_2$ &  51.4(1.7)  & 20.05(9)  \\
		$T_3$ &  32.8(1.0)  & 19.90(9)  \\
		$T_4$ &  20.6(6)  & 19.93(10)  \\   
		$T_5$ &  12.8(4)  & 19.87(11) \\
		$T_6$ &  8.03(22)   & 19.75(12)  \\
		$T_7$ &  4.91(13)   & 19.74(15) \\
		$T_8$ &  3.04(8)  & 19.58(17) \\
		\hline
	\end{tabular}
    \caption{Continuum extrapolated results for the normalized entropy density 
    $s/T^3$ at the 9 temperatures considered in this study.}
    \label{tab:entropy_cont}
\end{table}

We performed several checks to further corroborate the robustness of our 
best fit.
The continuum results are stable against adding a
term like $ (a/L_0)^2 g_i^4$ to Eq.~(\ref{eq:fit_type}).
Furthermore, we repeated the whole analysis replacing $d_3(a/L_0)^3g_i^3$
with $d_4(a/L_0)^4g_i^3$ in the fit anzatz. 
We obtain perfectly compatible results for the $s(T_i)$ with errors that are 
$10$-$20\%$ smaller. 
Similar conclusions hold when replacing $g_i^3\to g_i^4$ in 
both $a^2$ and $a^3$ terms.
Finally, we also checked the impact of logarithmic corrections to the
leading discretization effects, of O($a^2$), using the modified fit 
function~\cite{Husung:2019ytz,Husung:2021mfl,Husung:2022kvi}
\begin{equation}
    s(T_i, a/L_0)/T_i^3 = 
    s(T_i)/T_i + d_2 \left[\bar g_{\rm SF}^2(\pi/a)\right]^\gamma
    \left(\frac{a}{L_0}\right)^2 g_i^2\,.
\end{equation}
The results for $s(T_i)$ change by less than $1$ standard deviation with respect
to the  best fit, $\gamma=0$, when the effective anomalous dimension is 
varied in the interval $\gamma\in[-1,1]$. The comparison for three selected 
values is shown in the right panel of Figure~\ref{fig:clim}.

\section{Discussion}
\begin{figure}
     \centering
     \begin{subfigure}{0.48\textwidth}
         \centering
         \includegraphics[width=\textwidth]{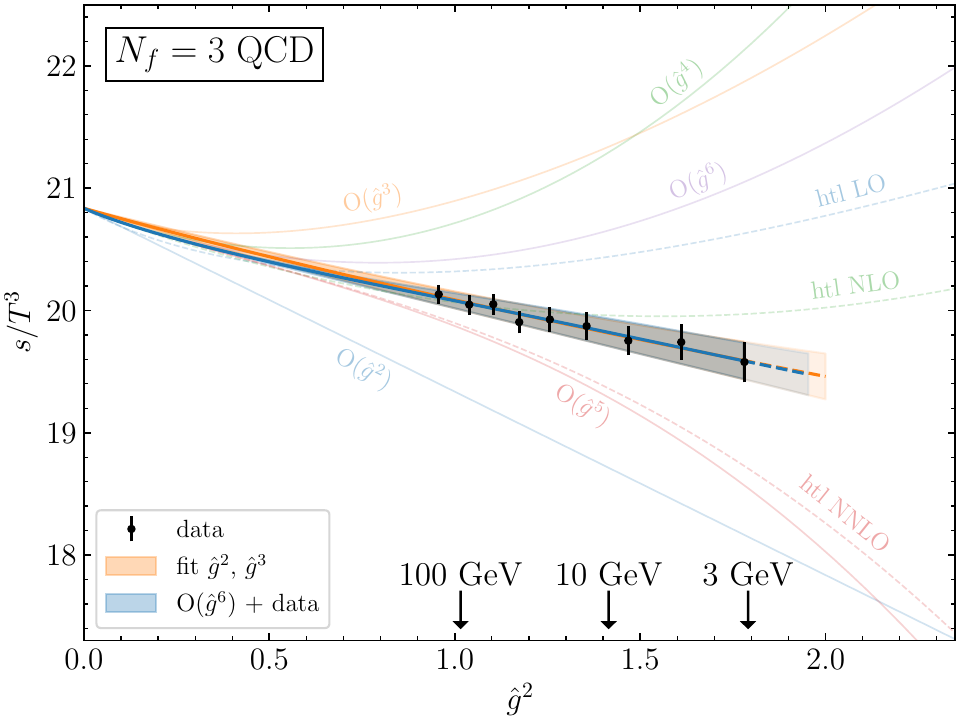}
     \end{subfigure}%
     \hspace{0.2cm}
     \begin{subfigure}{0.48\textwidth}
         \centering
         \includegraphics[width=\textwidth]{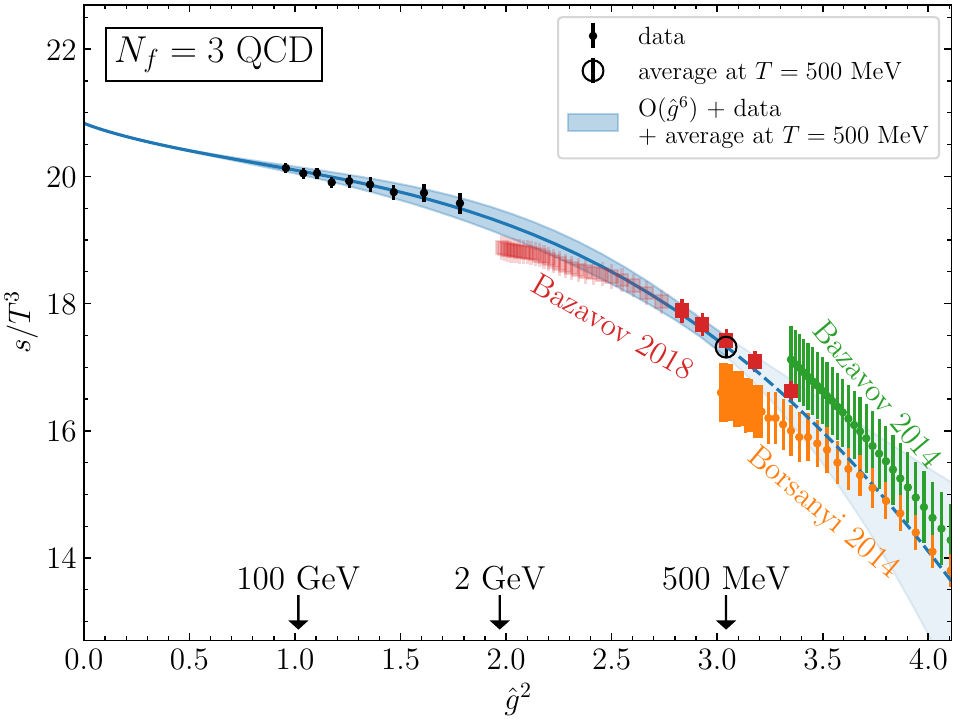}
     \end{subfigure}
     \caption{Left: Normalized entropy density $s/T^3$ in the continuum limit 
     as a function of $\hat g^2(T)$. The shadowed curves represent the 
     perturbative value up to the indicated order, and the prediction from
     hard-thermal-loop (htl)~\cite{Andersen:2010ct,Andersen:2011sf}. 
     The orange and blue bands (grey when the two overlap) correspond to 
     the fits to Eqs.~\eqref{eq:fitpheno} and~\eqref{eq:fitmikko}, see main text
     for the details.
     Right: $s/T^3$ as a function of $\hat g^2(T)$. The black points are
     our continuum values. At lower temperatures results from the literature
     are also shown, labelled as ``Borsanyi 2014''~\cite{Borsanyi:2013bia}, 
     ``Bazavov 2014''~\cite{HotQCD:2014kol} and 
     ``Bazavov 2018''~\cite{Bazavov:2017dsy}. 
     The black circle marker is the weighted 
     average of these results at $T=500$ MeV. Finally, the blue band is our
     best parametrization of $s/T^3$ for $N_f=3$ QCD at $T\geq 500$ MeV.}
     \label{fig:final_results}
\end{figure}
We parametrize the temperature dependence of the continuum extrapolated entropy density through the 
function $\hat g^2(T)$, defined as
the 5-loop strong coupling in the $\overline{\rm MS}$ scheme~\cite{Baikov:2016tgj} renormalized at the scale $2\pi T$. 
Its leading order expression reads
\begin{equation}
    \frac{1}{\hat g^2(T)} = \frac{9}{8\pi^2}\ln\frac{2\pi T}{\Lambda_{\overline{\rm MS}}} + \cdots\,,
\end{equation}
where the value of the Lambda parameter is $\Lambda_{\overline{\rm MS}} = 341$ MeV~\cite{Bruno:2017gxd}.
For our purposes this is only a convenient choice which makes simpler to compare 
the non-perturbative results with the prediction of perturbation theory.
In left panel of Figure~\ref{fig:final_results} we plot the continuum results 
of the entropy density reported in Table~\ref{tab:entropy_cont} as a function of $\hat g^2$.

A simple parametrization is a polynomial in $\hat g$ of the form
\begin{equation}
    \frac{s}{T^3} = \frac{32\pi^2}{45}\left[s_0 + s_2 \left(\frac{\hat g}{2\pi}\right)^2 
    + s_3 \left(\frac{\hat g}{2\pi}\right)^3\right]\,.
    \label{eq:fitpheno}
\end{equation}
We first set $s_3=0$ taking $s_0$ and $s_2$ as fit parameters. The resulting values are $s_0 =
2.954(15)$ and $s_2 = -3.6(7)$: notably, the former is in agreement within 1 standard deviation with the Stefan-Boltzmann (SB) value
$s_0^{\rm SB} = 2.969$, obtained in the infinite temperature limit. We thus fit our results enforcing $s_0=s_0^{\rm SB}$ in
Eq.~\eqref{eq:fitpheno} and leaving $s_2$, $s_3$ as fit parameters. The resulting coefficients are $s_2=-5.1(9)$, $s_3=5(5)$ and this fit is
represented as the orange band in the left panel of Figure~\ref{fig:final_results}. 
Thus, a simple, third-order polynomial in $\hat g$ is sufficient for the description of the entropy density
in the interval of temperatures considered in this study, which spans almost two orders of magnitude. However, we also point out that 
the slope in $\hat g^2$ resulting from that fit differs from the perturbative prediction $-8.438$ by a few standard 
deviations.

It is interesting to compare our non-perturbative results with the expectations coming from perturbative computations performed at high
temperatures. We thus consider a fit function with all the known perturbative coefficients,
\begin{equation}
    \frac{s}{T^3} = \frac{32\pi^2}{45}\left[
    \sum_{k=0}^6 s_k \left(\frac{\hat g}{2\pi}\right)^k
    + q_c \left(\frac{\hat g}{2\pi}\right)^6 
    + s_7 \left(\frac{\hat g}{2\pi}\right)^7\right]\,,
    \label{eq:fitmikko}
\end{equation}
where $s_0,...,s_6$ are obtained from the perturbative expansion of the pressure~\cite{Kajantie:2002wa}.
The fit function~\eqref{eq:fitmikko} is thus expected to reproduce the correct behaviour at asymptotically high temperatures.
We have two fit parameters: $q_c$ represents the unknown contributions at order O$(\hat g^6)$ including those from the 
non-perturbative ultrasoft modes, and $s_7$ takes into account higher order effects in the weak coupling expansion.
By fitting our data we obtain $q_c=-5.1(1.7)\cdot10^3$, $s_7=1.3(7)\cdot10^4$ 
with $\chi^2/\chi^2_{\rm exp}=0.58$. The corresponding curve is displayed in blue in the left panel of Figure~\ref{fig:final_results}. 
The two fits Eq.~\eqref{eq:fitpheno} with the SB value enforced, and Eq.~\eqref{eq:fitmikko} are in very good agreement in the 
interval of temperatures considered in this study, and also for $T\to\infty$.

For temperatures up to about $500$ MeV, the EoS of $N_f=2+1$ QCD has been 
computed in the continuum limit by the Wuppertal-Budapest 
collaboration~\cite{Borsanyi:2013bia} and by the HotQCD collaboration
~\cite{HotQCD:2014kol,Bazavov:2017dsy}.
Since the contribution of the light quark masses at $T=500$ MeV is negligible
within the statistical accuracy quoted by the two collaborations~\cite{Laine:2006cp}, 
we can compare these results with ours. 

Including in the fit of Eq.~\eqref{eq:fitmikko} the value $s/T^3 = 17.31(16)$ at $T=500$ MeV,
obtained from the average of the results of the two collaborations, the resulting fit parameters are $q_c = -4.0(1.1)\cdot 10^3$, $s_7 =
7(4)\cdot 10^3$ with $\chi^2/\chi^2_{\rm exp}=0.79$. This is our best parametrization of the normalized entropy density of
$N_f=3$ QCD for $T\geq 500$ MeV, and it is represented as the blue shadowed
curve in the right panel of Figure~\ref{fig:final_results}.
The relative error is $\lesssim 1\%$ in the whole temperature range considered.

Following Ref.~\cite{Kajantie:2002wa}, the pressure can be parameterized analogously to Eq.~\eqref{eq:fitmikko} with
$p_7=s_7 + (5\, b_0 p_5 + 3\, b_1 p_3)/4$, where $b_0=9/4$, $b_1=4$, and $p_3$ and $p_5$ are given in Ref.~\cite{Kajantie:2002wa}. The energy
density is then determined using the relation $e=Ts-p$.

\section{Conclusions}

In a recent publication~\cite{Bresciani:2025vxw}, and in these Proceedings, 
we show the first non-perturbative computation of the EoS of QCD with 
$N_f=3$ massless flavours, in the range of temperatures from $3$ to $165$ GeV.
The final accuracy on our primary observable, the entropy density,
is $0.5$-$1.0\%$ and is dominated by the statistical error.
By combining our data with non-perturbative results at $T=500$ MeV from the 
literature and by enforcing the behaviour at asymptotically high
temperatures using the known perturbative expansion, we could determine 
the temperature dependence of the entropy density for $T\geq 500$ MeV with
a relative error of at most $\sim 1\%$.
The pressure and the energy density can then be computed
as well using standard thermodynamic relations~\cite{Bresciani:2025vxw}.

This result relies on a completely new strategy, based on two pillars: on the one hand, we determine
the lines of constant physics at very high temperature by imposing
the renormalization conditions through the non-perturbatively defined 
strong coupling $\bar g_{\rm SF}^2$ of QCD, whose running in the continuum 
is known with high accuracy~\cite{Bruno:2017gxd}.
This gives us the freedom to choose the renormalization scales, at which the 
bare parameters are renormalized, to be close to the temperatures 
simulated~\cite{DallaBrida:2021ddx}.
On the other hand, we formulate QCD in a moving reference frame~\cite{Giusti:2010bb,Giusti:2012yj} which represents 
a convenient setup where thermodynamic potentials can be determined without the need of any zero-temperature subtraction.

Our strategy can be readily generalized to the case
of QCD with four or five (massive) flavours. This result would directly 
improve the present knowledge of the Standard Model EoS, whose uncertainty
for temperatures from a few GeV up to the electro-weak scale is 
dominated by the systematics introduced by the poor convergence of the 
perturbative expansion of the QCD component~\cite{Saikawa:2018rcs,Saikawa:2020swg}.

Finally, we also notice that in the presence of shifted boundary 
conditions the EoS can be accessed through suitable one-point functions
of the Energy-Momentum tensor (EMT)~\cite{Giusti:2012yj}.
On the lattice, this gives us the handle to compute in a non-perturbative way 
the renormalization constants that define the lattice EMT.
Some preliminary results for the renormalization factors of the non-singlet
components with a precision of a few percent are reported in Ref.~\cite{Bresciani:2023zyg}.
This is a notoriously challenging theoretical problem in 
lattice QCD~\cite{Caracciolo:1989pt,Caracciolo:1989bu,Caracciolo:1991vc,Caracciolo:1991cp}, 
whose solution would open the way to further first-principle investigations of the thermal properties of QCD 
through the correlation functions of the (renormalized) EMT~\cite{Meyer:2011gj}.

\acknowledgments
We acknowledge PRACE for awarding us access to the HPC system MareNostrum4 
at the Barcelona Supercomputing Center (Proposals n. 2018194651 and 2021240051)
where some of the numerical results presented in this letter have been obtained. 
We also thank CINECA for providing us with a very generous access to Leonardo 
during the early phases of operations of the machine and for the computer time 
allocated via the CINECA-INFN, CINECA-Bicocca agreements. 
The R\&D has been carried out on the PC clusters Wilson and Knuth at 
Milano-Bicocca. 
We thank all these institutions for the technical support. 
This work is (partially) supported by ICSC – Centro Nazionale di Ricerca in High
Performance Computing, Big Data and Quantum Computing, funded by European 
Union – NextGenerationEU.

\bibliographystyle{JHEP}
\bibliography{bibfile}

%\begin{thebibliography}{99}
%\input{bibliography}
%\end{thebibliography}

\end{document}

%% file: include.tex
\usepackage{graphicx}
\usepackage{graphics}
\usepackage{multirow}
\usepackage{multicol}
\usepackage{caption}
\usepackage{subcaption}

\usepackage{amsfonts}
\usepackage{amsmath}
\usepackage{mathtools}
\DeclareMathAlphabet{\mathcal}{OMS}{cmsy}{m}{n}

\newcommand{\Nf}{{N_f}}
\newcommand{\psibar}{{\overline{\psi}}}

\newcommand{\cZ}{{\cal Z}}

\newcommand{\corrno}[1]{{\langle#1\rangle}}
\newcommand{\bs}[1]{{\boldsymbol{#1}}}

\newcommand{\vxi}{{\bs \xi}}

\usepackage{xcolor}